\documentclass[aps,prb,onecolumn,,floatfix]{revtex4-2}

\usepackage{amsmath}
\usepackage{amssymb}
\usepackage{graphicx, float}
\usepackage{algorithm}
\usepackage{algpseudocode}

\newcommand{\bea}{\begin{eqnarray}}
\newcommand{\eea}{\end{eqnarray}}

\newcommand{\parent}[1]{\left( #1 \right)}
\newcommand{\mean}[1]{\left \langle #1 \right \rangle}

\begin{document}


\title{Exact method for calculating the current fluctuations and nonlinear response of Markovian dynamics}

\author{David Andrieux}

\noaffiliation


\begin{abstract}
We show that the current fluctuations and nonlinear response of Markovian dynamics can be obtained from a system of polynomial equations.
This offers new opportunities for analytical and numerical results. 
As an example, we derive new expressions for the current nonlinear response coefficients.
\end{abstract}


\maketitle

\section{Context and objectives}

Current fluctuations and their nonlinear response capture important dynamical and thermodynamical properties of stochastic systems.
Both fluctuations and their response are closely intertwined, as revealed by the fluctuation theorem for currents \cite{AG04}.
More recently, we showed that the nonlinear response takes a simple, fully symmetric form when the system parameters are varied along dynamical equivalence classes \cite{A12b, A12c, A22}

However, calculating current fluctutations and these dynamical equivalence classes remain difficult.
In the traditional approach, it requires solving an eigenvalue and eigenvector problem for each value of the counting parameters. 
While this can usually be done numerically efficiently, it is difficult to gain additional insights using this approach.

In this paper we formulate the calculation of the current generating function and the associated nonlinear response as a system of polynomial (cubic) equations.
This method is based on a decomposition of Markov dynamics into cycles \cite{C81, A83, K94} combined with a partition into dynamical equivalence classes \cite{A12b, A12c}. 
This approach offers an alternative way to numerically evaluate these quantities, as well as opportunities to derive new analytical insights.
As an example, we use this formulation to derive new expressions for the nonlinear response coefficients.


\section{Markov chains and decomposition in cycle matrices}

We consider a Markov chain characterized by a transition matrix $P = \parent{P_{ij}} \in \mathbb{R}^{N\times N}$ on a finite state space.
We assume that the Markov chain is primitive, i.e. there exists an $n_0$ such that $P^{n_0}$ has all positive entries. 
The chain $P$ thus admits a unique stationary distribution $\pi$.

It will be convenient to refer to the associated matrix $F$ such that $F_{ij} = \pi_i P_{ij}$. 
Each element $F_{ij}$ corresponds to the steady state probability flux between state $i$ and $j$.
$F$ satisfies 
\bea
\sum_i F_{ij} = \pi_j \quad \text{and} \quad \sum_j F_{ij} = \pi_i
\label{ss}
\eea
so that $\sum_{ij} F_{ij} =1$. Every matrix $F$ corresponds to a unique chain $P$ and vice versa. 

Cohen \cite{C81} and Alpern \cite{A83} demonstrated that the matrix $F$ can always be decomposed as a combination of {\it cycle matrices}. 
If $[a_1, \cdots, a_m]$ is a sequence of distinct integers chosen from $1, \cdots, N$, then we define the corresponding cycle matrix as the $N \times N$ matrix $C^{[a_1, \cdots, a_m]}$ given by $c_{a_1a_2} = c_{a_2a_3} = \cdots = c_{a_m a_1} = 1/m$ and $0$ otherwise.  
We say that $\ell = m$ is the length of $C$.
Then, for some vector $(\lambda_1, \cdots, \lambda_l)$ such that $\sum_e \lambda_e =1$ and $\lambda_e \geq 0$, and some cycle matrices $C^e$, we have 
\bea
F = \sum_e \lambda_e \, C^e \, .
\label{decomp}
\eea
Note that the decomposition (\ref{decomp}) is not unique. 
In this paper, we will use a decomposition of the form (\ref{decomp}) with a clear thermodynamic interpretation \cite{S76, K94, K06}. 

We represent the chain $P$ by a graph where each vertex corresponds to a state and each transition $P_{ij} > 0$ by an edge $e$. 
Then, for a graph with $N$ vertices and $E$ edges with $i \neq j$ and $D$ edges with $i = j$, there exists $M = E-N+1$ independent thermodynamic currents $J_\alpha$, with each current defined by a fundamental cycle $C^\alpha$ (see Ref. \cite{S76} for details). 
We then have the following result:\\

{\bf Theorem}: Denote by $e$ the set of cycles $[i,j]$ with $P_{ij}>0$ (and thus $P_{ji} >0$ since P is primitive) and by $\alpha$ the fundamental cycles. 
Every matrix $F$ is a convex combination of $D+E+M = D+ 2E - N +1 \leq 2E -1$ cycle matrices
\bea
F = \sum_e \lambda_e \, C^e + \sum_\alpha \lambda_\alpha \, C^\alpha\, .
\label{Tdecomp}
\eea
with $\sum_e \lambda_e + \sum_\alpha \lambda_\alpha =1$ and $(\lambda_e, \lambda_\alpha) \geq 0$. \\

{\it Demonstration}: Combine results from Refs. \cite{S76, K94, K06, A23}.\\

The representation (\ref{Tdecomp}) separates the nonequilibrium components $\lambda_\alpha = J_\alpha \ell_\alpha$ from the 'local' fluxes $\lambda_e$ that do not contribute to the net currents.
In particular, $F$ is symmetric and thus corresponds to an equilibrium dynamics if and only if $\lambda_\alpha = 0$ for all $\alpha$.

We can also obtain the time reversal of a Markov chain from the decomposition (\ref{Tdecomp}). 
The time reversal of a Markov chain is defined as
\bea
P^R = {\rm diag}(\pi)^{-1} \, P^T \, {\rm diag}(\pi)
\eea
where $\pi$ is the stationary distribution of $P$. Using (\ref{Tdecomp}) and noting that $(C^e)^T = C^e$ and $(C^\alpha)^T = C^{\alpha^R}$ where $\alpha^R$ is the reverse cycle $[a_m, \cdots, a_1]$, we obtain that the associated flux matrix
\bea
F^R = \sum_e \lambda_e C^e + \sum_\alpha \lambda_\alpha C^{\alpha^R} \, .
\label{FTR}
\eea
The time reversed chain $P^R$ therefore has the same thermodynamic currents as $P$ but in the reverse directions.
We will use this observation to reduce the size of computation of the current fluctuations in half. \\

{\bf Example: Transport on a disordered ring.} Throughout the paper, we will illustrate our results by looking at a disordered ring. 
We consider the Markov chain $P$ representing a ring of length $\ell = N$ with periodic boundary condition and no self-transition: $P_{ij} > 0$ if $|i-j|=1$ or if $|i-j|=\ell-1$, and $0$ otherwise. 
In this case $D=0$ and $E=N$ so that the system possesses one independent current $J_\alpha = J$ associated with the cycle $\alpha = [1,2, \ldots, \ell]$ if $J>0$ and $[\ell, \ldots, 2, 1]$ if $J<0$.
If $k$ denotes the cycles $[k, k+1]$ (with periodic boundary conditions), we have that any such ring dynamics can be written as
\bea
F= \lambda_{[1,2,..., \ell]} C^{[1,2,..., \ell]} + \sum_{k} \lambda_{[k,k+1]} C^{[k,k+1]} \nonumber 
\label{perring}
\eea
where $\lambda_{[1,2,..., \ell]} = J \ell$ and
\bea
 C^{[1,2,..., \ell]} =
\frac{1}{\ell}
  \begin{pmatrix}
    0 &  1 & 0 & \cdots & 0\\
    0 & 0 & 1 & \ddots & 0 \\
    \vdots &  & \ddots & \ddots & 1 \\
    1 & 0 & \cdots & 0 & 0 \\
  \end{pmatrix} 
\, ,
\quad \quad 
 C^{[k, k+1]} =
\frac{1}{2}
  \begin{pmatrix}
   \ddots & & & &   \\
      &  & 0 & 1 &  \\
     &  & 1 & 0 & \\
      &  &   &  & \ddots \\
  \end{pmatrix} 
\nonumber
\eea 
where $(C^{[k,k+1]})_{k, k+1} = (C^{[k,k+1]})_{k+1, k} = 1$ and $0$ otherwise. 
A similar decomposition applies when $J<0$ with $C^{[1, 2, , ...,\ell]}$ replaced by $C^{[\ell, ..., 2,1]}$, i.e. the case $J<0$ corresponds to the time reversed chain. 

The steady state probabilities are readily obtained through Eqs (\ref{ss}), giving $\pi_i = \lambda_{[1,2,..., \ell]}/ \ell + \lambda_{i, i+1}/2 + \lambda_{i-1, i}/2.$


\section{Current fluctutations and generating function}

The Markov chain $P$ generates random trajectories $i_0 \rightarrow i_1 \rightarrow \ldots \rightarrow i_n$.
The fluctuating currents are then measured by
\bea
G_\alpha (n) =\sum_{l=1}^n j_\alpha (l) \, ,
\eea
where
$j_\alpha (l) = \pm 1$ if the transition $i_{l-1}\rightarrow i_{l}$ corresponds to the fundamental edge $\alpha$ in the positive (negative) direction, and $0$ otherwise \cite{S76, AG07}.
 
The fluctuations of the currents $j_\alpha$ can be characterized by the cumulant generating function
\bea
q (\pmb{s}) = \lim_{n\rightarrow\infty} \frac{1}{n} \ln \mean{ \exp \parent{ {\sum_\alpha s_\alpha G_\alpha (n)} }  } \, .
\label{q}
\eea
All the cumulants are obtained by successive derivations with respect to the counting parameters $\pmb{s}$.

The cumulant generating function (\ref{q}) is given by \cite{AG07, A22}
\bea
q (\pmb{s}) = \ln \rho [ P \circ Z ( \pmb{s} ) ] \, ,
\label{rho}
\eea
where $\rho (Y)$ denotes the spectral radius (i.e. the modulus of the largest eigenvalue) of the operator $Y$, $\circ$ is the Hadamar product of two operators, and
\bea
Z_{ij} (\pmb{s}) \equiv \begin{cases} \exp \parent{+ s_\alpha}  & \text{if the transition $i\rightarrow j$ corresponds to the edge $\alpha$ in the positive direction,} \\
                 \exp \parent{-s_\alpha}  & \text{if the transition $i\rightarrow j$ corresponds to the edge $\alpha$ in the negative direction,} \\
                1 & \text{otherwise}.
                \end{cases}
\nonumber
\eea
The operator $P\circ Z$ is non-stochastic when $\pmb{s} \neq 0$.
Therefore, its spectral radius, and thus the generating function, often cannot be resolved analytically.
Obtaining the generating function then requires calculating the Perron eigenvalue of the operator $P\circ Z$. 
Note that, as $|\pmb{s}|$ increases, some elements of $P\circ Z$ become exponentially large while other become exponentially small, making the numeral evaluation of $q$ difficult.

The Perron eigenvectors $x_{\pmb{s}}$ define a path in the space of Markov dynamics through
\bea
P^*(\pmb{s}) = \frac{1}{\rho(\pmb{s})} {\rm diag}(x_{\pmb{s}})^{-1}\, [P\circ Z(\pmb{s})]\, {\rm diag}(x_{\pmb{s}}) \, .
\label{Pstar}
\eea
These dynamics play a special role in the nonlinear response theory \cite{A12b, A12c}. 
In particular, the counting parameters $\pmb{s}$ correspond to the affinities of these dynamics: $\pmb{A}[P^*] = 2\pmb{s}$.
We will return to these observations in the next sections.\\

{\bf Example (continued).} For the disordered ring, the single independent current can be measured along any edge $i \rightarrow i+1$. 
Choosing the fundamental transition as $\alpha = \ell \rightarrow 1$ the current generating function is given by the largest eigenvalues of 
\bea
P \circ Z (s) =
  \begin{pmatrix}
    0 &  P_{1,2} & 0 & \cdots & P_{1,\ell} \, {\rm e}^{-s}\\
    P_{21} & 0 & P_{23} & \ddots & 0 \\
    \vdots &  & \ddots & \ddots & P_{\ell-1, \ell} \\
    P_{\ell,1} \, {\rm e}^{+s} & 0 & \cdots & P_{\ell, \ell-1} & 0 \\
  \end{pmatrix}   \, .
\nonumber
\eea
The affinity of the resulting dynamics (\ref{Pstar}) takes the value $A=2 s$.


\section{Dynamical equivalence classes and alternative method to calculate the current fluctuations}

In this section we derive an alternative method to calculate the current generating function (\ref{q}).
To this end, we build on the fact that the dynamics (\ref{Pstar}) are related through the relation
\bea
P^* \circ (P^*)^T (\pmb{s}) = \gamma (\pmb{s},\pmb{s}') \, P^* \circ (P^*)^T (\pmb{s}')
\label{Pgamma}
\eea
where $\gamma = \rho^2 (\pmb{s}')/ \rho^2 (\pmb{s})$. This relation defines equivalence classes in the space of Markov dynamics \cite{A12b}.
In Ref. \cite{A12b} we also proved that the factor $\gamma$ is related to the current generating function by $q = - (1/2) \log \gamma + C$. 

We now consider the equivalence class $[P] = \{ P^*(\pmb{s}) \}$ and how to generate it using the decomposition (\ref{Tdecomp}).\\ 

{\bf Theorem}: Let $E = P\circ P^T$. 
The equivalence class $[P]$ is parametrized by the independent currents $\lambda_\alpha = J_\alpha \ell_\alpha$. 
In particular the associated dynamics $F(\lambda_\alpha) \in [P]$ satisfy
\bea
F_{ij} F_{ji} = \gamma \, E_{ij} \, \Big( \sum_j F_{ij} \Big) \Big( \sum_i F_{ij} \Big)
\label{fullpol}
\eea
for some $\gamma$.\\

{\it Demonstration}: The equivalence class (\ref{Pgamma}) is defined by $[P] = \{ T | \, T_{ij}T_{ji} = \gamma E_{ij}\}$ for some $\gamma$.
Inserting the definition $F_{ij} = \pi_i T_{ij}$ and using that $\sum_j F_{ij} = \pi_i$ we obtain equation (\ref{fullpol}). \\

This result can now be used to obtain an alternative method to calculate the current fluctuations.
For simplicity we start from an equilibrium dynamics $\bar{P}$, the method can be directly extended to cover the case where the starting dynamics is out of equilibrium.
In addition, we showed that the generating function of an arbitrary nonequilibrium dynamics can be obtained as the translation of an equilibrium generating function \cite{A12c}.
\\

\noindent
{\bf Alternative method for the calculation of current fluctuations}\\

Let $\bar{P}$ be an equilibrium dynamics, $\bar{F} = \sum_e \bar{\lambda}_e C^e$ its associated flux matrix, and $E = \bar{P}\circ \bar{P}^T$. 
We will vary the thermodynamic currents $\lambda_\alpha = J_\alpha \ell_\alpha \in [0,1]$. 
For each value of these currents, we consider the $D+E+1$ variables $(\pmb{\lambda}_e, \gamma)$.

According to (\ref{fullpol}), these $D+E+1$ variables satisfy the system of $D+E+1$ polynomial (cubic) equations
\bea
\begin{cases} 
[ \lambda_e/2 +\sum_\alpha \max(\epsilon_\alpha (e),0) \lambda_\alpha/\ell_\alpha ] [ \lambda_e/2 - \sum_\alpha \min(\epsilon_\alpha (e),0) \lambda_\alpha/\ell_\alpha ] = \gamma \, E_{ij} \pi(i) \pi (j) \\
\sum_e \lambda_e + \sum_\alpha \lambda_\alpha = 1 \, ,
\end{cases} 
\label{explicitpol}
\eea
where for notation simplicity we introduced $\pi(i) = \sum_j F_{ij} = \sum_j [\sum_e \lambda_e (C^e)_{ij} + \sum_\alpha \lambda_\alpha (C^\alpha)_{ij}]$.
Here the $\max / \min$ functions select the positive and negative currents along edge $e$, respectively. 
While the equations (\ref{explicitpol}) look complicated, in practice they simplify considerably for a given network topology. 
Also, they can easily be implemented for numerical evaluation \cite{FN05}. 

The generating function for the dynamic $F(\pmb{\lambda}) = F(\pmb{\lambda}_e, \pmb{\lambda}_\alpha)$ is then given by $q = - (1/2) \ln \gamma (\pmb{\lambda})$. 
We can then solve this system for multiple values of the currents $\lambda_\alpha = J_\alpha \ell_\alpha$ to obtain the full generating function. 
In particular, at equilibrium $\pmb{\lambda}_\alpha = \pmb{J}_\alpha = 0$ and the solution takes the form $(\bar{\lambda}_e, 1)$. 

To obtain the parametrization of the generating function in terms of the counting parameters $\pmb{s}$, we simply use the relation $2 \pmb{s} = \pmb{A}(\pmb{J})$ to obtain $q = - (1/2) \ln \gamma [\pmb{A}(\pmb{J})/2]$.\\

\begin{figure}[h]
\includegraphics[scale=.55]{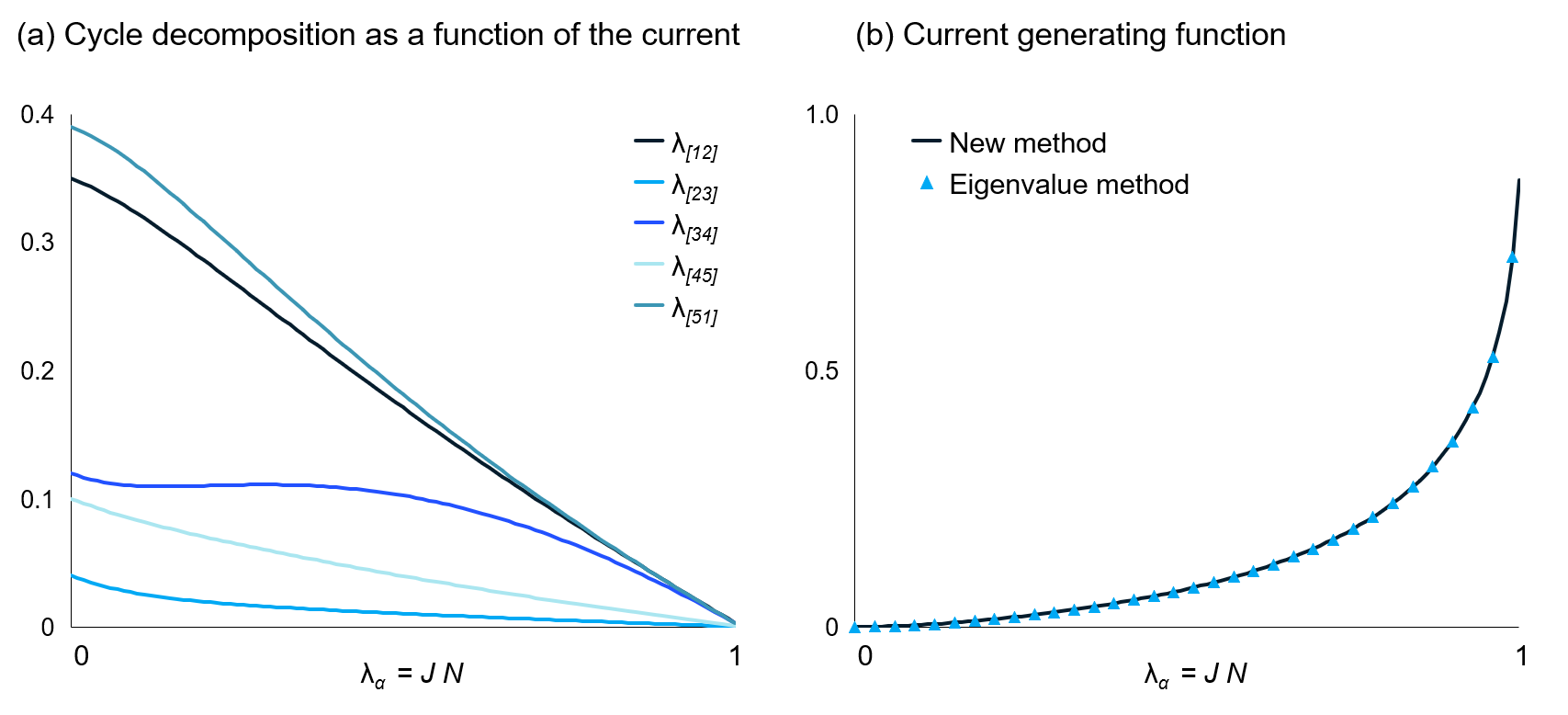}%
\caption{{\bf Transport on a disordered ring}. 
(a) Numerical evaluation of the system (\ref{ringpol}) as a function of $\lambda_\alpha = J \ell$. 
At equilibrium $\gamma=1$ and decreases to $\gamma = 0$ at maximal (irreversible) current $\lambda_\alpha = 1$. 
(b) Corresponding current generating function, calculated as $q = -(1/2) \ln \gamma$ with the method (\ref{ringpol}) (solid line) and the eigenvalue problem (\ref{rho}) (triangles). 
By the fluctuation theorem the generating function is symmetric (not shown); the dynamics for negative values of the current are obtained by the time-reversed dynamics (\ref{FTR}).
The parameters take the value $N=\ell = 5$, and the equilibrium dynamics is defined by $\lambda_{[1,2]} =0.35, \lambda_{[2,3]} =0.04, \lambda_{[3,4]} =0.12, \lambda_{[4,5]} =0.1, \lambda_{[5,1]} =0.39, \gamma =1$.}
\label{Ring}
\end{figure}

In other words, we can solve the Perron eigenvalue and eigenvector problem by solving this polynomial system of $D+E+1$ equations.
One advantage of this procedure is that all variables are bounded, $0 \leq \lambda_e \leq 1$ and $0 \leq \gamma \leq 1$ (Fig. \ref{Ring}). 
In this way the system of equation is well-behaved even for large $\pmb{s}$, which corresponds to $\gamma \rightarrow 0$ and $\lambda_k \rightarrow 0$.\\

{\bf Example (continued).} For the disordered ring, the system of $\ell + 1$ equations (\ref{explicitpol}) takes the form
\bea
\begin{cases} 
\lambda_k (\lambda_k + 2\lambda_\alpha/\ell) = \gamma \, E_k (\lambda_k +\lambda_{k-1}  + 2\lambda_\alpha/\ell) (\lambda_k +\lambda_{k+1}  + 2\lambda_\alpha/\ell)  \quad {\rm for} \quad k=1,...,\ell \\
\sum_k \lambda_k + \lambda_\alpha = 1
\end{cases} 
\label{ringpol}
\eea
where we used that $\pi_k = \lambda_k/2 +\lambda_{k-1}/2  + \lambda_\alpha/\ell$. 
The current generating function calculated based on this system of $\ell+1$ variables is illustrated in Fig. (\ref{Ring}). 

\section{Expressions for the nonlinear response coefficients}

\begin{figure}[h]
\includegraphics[scale=.55]{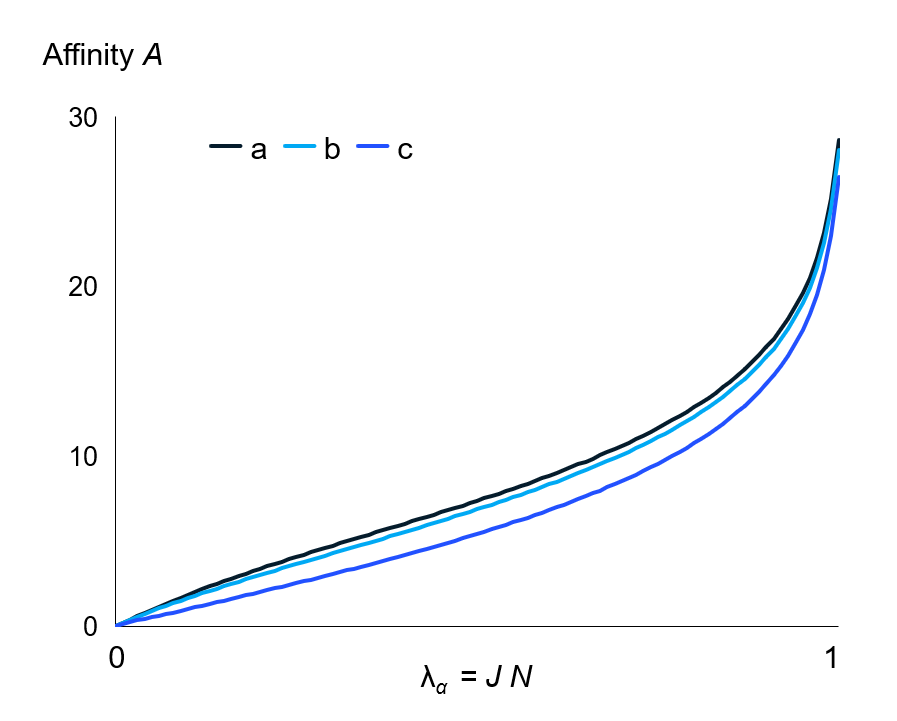}
\caption{ {\bf Nonlinear response curves for different variations of parameters as a function of the current.} 
(a) Intrinsic dynamics (\ref{Pstar}) giving rise to the reponse curve (\ref{L.b}).
(b) Linear decrease $\lambda_e (\lambda_\alpha) = \bar{\lambda}_e (1- \lambda_\alpha)$.
(c) Quadratic interpolation between $\bar{\lambda}_e$ and $\lambda_e (\lambda_\alpha') = (1-\lambda_\alpha')/N$ for $\lambda_\alpha \leq \lambda_\alpha' =0.1$ and $\lambda_e (\lambda_\alpha') = (1-\lambda_\alpha)/N$ for $\lambda_\alpha \geq \lambda_\alpha'$.
Note that for some parameters variations the affinity diverges for current values $\lambda_\alpha < 1$. 
For example, for the path $\lambda_e (\lambda_\alpha) = \bar{\lambda}_e$ for all $e$ except $\lambda_1 (\lambda_\alpha) = \bar{\lambda}_1 - \lambda_\alpha$ the dynamics become irreversible at $\lambda_\alpha =\bar{\lambda}_1$ and the affinity diverges at that point (not shown).
The equilibrium parameters $\bar{\lambda}_e$ take the same values as in Figure (\ref{Ring}).}
\label{Fig2}
\end{figure}

The current response theory expands the currents as functions of the affinities around equilibrium ($\pmb{A} =0)$:
\bea
J_\alpha = \sum_{l=1}^{\infty} \frac{1}{l!} L_{\alpha, \beta_1 ... \beta_l} \, A_{\beta_1} \cdot\cdot\cdot \ A_{\beta_l} \, ,
\label{Jexp}
\eea
where we sum over repeated indices. 
The linear response $L_{\alpha, \beta}$ can be expressed as equilibrium correlations \cite{AG04, H05}. 
Further away from equilibrium however, the response coefficients depend on how the system parameters are varied are not thus not uniquely defined \cite{H05}. 
This can be easily understood since there are $M = N-E+1$ independent affinities while there are $D+E+M > M$ independent transition probabilities that define a Markov chain $P$ with the same network topology.
Therefore, there are multiple dynamics that can achieve a set of affinities $\pmb{A}$ (Figure \ref{Fig2}). 

By varying the affinities along the equivalence class (\ref{Pgamma}) the nonlinear response coefficients take the simple form \cite{A22}
\bea
L_{\alpha, \beta_1 ... \beta_l} &=& 0  \quad  \quad \quad \quad \quad \quad \quad  \quad \text{if {\it l} is even} \label{L.a} \\
L_{\alpha, \beta_1 ... \beta_l} &=& \parent{\frac{1}{2}}^l K_{\alpha \beta_1 ... \beta_l} (\pmb{0}) \quad \text{if {\it l} is odd.}
\label{L.b}
\eea 
This shows that all response coefficients $L_{\alpha, \beta_1 ... \beta_l}$ are fully symmetric in ($\alpha, \beta_1, ..., \beta_l$). 
In particular, we recover the Onsager symmetry $L_{\alpha, \beta} = L_{\beta, \alpha}$ and the corresponding Green-Kubo formula, $L_{\alpha, \beta} = (1/2) K_{\alpha\beta} (\pmb{0}).$ 
In addition, all response coefficients are entirely expressed in terms of equilibrium correlations.

In view of these properties, we propose that this form constitutes the 'intrinsic' nonlinear reponse of a system.
By construction, the system (\ref{explicitpol}) generates the equivalence class (\ref{Pstar}). 
We can thus use it to calculate the response theory (\ref{L.a})-(\ref{L.b}).\\

In addition, we now show how the decomposition (\ref{Tdecomp}) provides analytical expressions for the response coefficients.
It will be easier to consider the series
\bea
A_\alpha = \sum_{l=1}^{\infty} \frac{1}{l!} B_{\alpha, \beta_1 ... \beta_l} \, J_{\beta_1} \cdot\cdot\cdot \ J_{\beta_l} \, ,
\label{Aseries}
\eea
which expresses the affinities in terms of the currents rather than the currents as function of the affinities as in (\ref{Jexp}). 
We can then reverse the series expansion to express the coefficients $L_{\alpha, \beta_1 ... \beta_l}$ in terms of the coefficients $B_{\alpha, \beta_1 ... \beta_l}$ \cite{FN01}.
 
We start by noting that the affinities $A_\alpha$ can be written as
\bea
A_\alpha = \sum_e A_e \epsilon_\alpha (e)
\label{Aalpha}
\eea
where $\epsilon_\alpha (e)$ takes the value $\pm 1$ if the edge $e$ belongs to $C^\alpha$ in the positive (negative) direction and $0$ otherwise \cite{S76}. 
The local affinities $A_e$ are in turn expressed in terms of the positive and negative fluxes along the edge:
\bea
A_{e} (\pmb{J}) = \ln \parent{ \frac{\lambda_e/2 +\sum_\beta \max[\epsilon_\beta (e),0] \lambda_\beta/\ell_\beta }{\lambda_e/2 - \sum_\beta \min[\epsilon_\beta (e),0] \lambda_\beta/\ell_\beta } } \, .
\label{Ae}
\eea
Importantly, the $\lambda_e$s are function of the currents, i.e. $\lambda_e = \lambda_e (\pmb{J})$, since $\sum_e \lambda_e + \sum_\alpha \lambda_\alpha  =1$.

We can then insert (\ref{Ae}) into (\ref{Aalpha}) and expand in power series of $J_\alpha = \lambda_\alpha/\ell_\alpha$ to express the affinties in terms of the currents.
After some algebra we obtain the linear response coefficients
\bea
B_{\alpha, \beta} = \sum_e \parent{ \frac{2}{\bar{\lambda}_e} } \epsilon_\alpha (e) \epsilon_\beta (e) \, .
\label{B.1} 
\eea
These coefficients depend on the equilibrium fluxes $\bar{\lambda}_e/2$ only, and are thus uniquely defined.
This formula was previously obtained by Hill \cite{H05} and Schnakenberg \cite{S76}.

The second-order coefficients read
\bea
B_{\alpha, \beta_1 \beta_2} = - \sum_e   \parent{ \frac{2}{\bar{\lambda}_e} }^2 \epsilon_\alpha (e) \parent{ \epsilon_{\beta_1} (e) \parent{\frac{\lambda'_{e, \beta_2}}{2}} -\frac{1}{2} u[\epsilon_{\beta_1} (e), \epsilon_{\beta_2}(e) ] }\, .
\label{B.2}
\eea
where $\lambda'_\beta = d\lambda/dJ_\beta$ and $u(a,b) = \max(a,0) \max(b,0) +\min(a,0) \min(b,0)$.
To the best of our knowledge, formula (\ref{B.2}) has not been derived before. 
We can derive similar expressions for the higher order coefficients using the same procedure, which will then depend on higher order variations $d^n\lambda/dJ^n$.

These results illustrates that the nonlinear response coefficients are not uniquely defined. 
Indeed, we see that the response coefficients depend on the path taken in the space of stochastic dynamics (here through $\lambda'$) and which can vary arbitrarily with the current $J$ (while respecting the constraint $\sum_e \lambda_e + \sum_\alpha \lambda_\alpha  =1$).
For example, if we follow the 'intrinsic' nonlinear response path defined by (\ref{Pstar}) the second order coefficient vanishes, $L_{\alpha, \beta_1\beta_2} = 0$, which translates into $B_{\alpha, \beta_1\beta_2} = 0$.\\

To recap, these results formulate the calculation of the current fluctuations, equivalence classes, and nonlinear response using a polynomial system of equations with bounded variables. 
This formulation is based on combining (1) a cycle decomposition of stochastic matrices, (2) a representation of these matrices in terms of thermodynamic currents, and (3) a partition of nonequilibrium reponse based on dynamical equivalence classes.
This opens the way for new analyses of the behavior of nonequilibrium systems, here starting with new expressions for the nonlinear response coefficients.

The decomposition (\ref{Tdecomp}) could also be combined with other methods to calculate the generating function. 
For example, it could be used to solve the minimization problem introduced by Shieh, which expresses the calculation of the spectral radius as the minimization of the Kublack-Leibler distance between two distributions \cite{S11, A12a}.

\vskip 1 cm

{\bf Disclaimer.} This paper is not intended for journal publication.


\begin{thebibliography}{99}
\bibitem{FN05} By construction the system admits a unique solution, although there doesn't seem to be a simple proof.
\end{thebibliography}
\end{document}